# Nepal's Engagement with the Millennium Challenge Corporation (MCC): A Philosophical and Economic Perspective

Daniel Loebell[1], Mingmar Sherpa[1], Ram Datta Bhatta[1]
[1]Department of Political Science, Northwestern University, Illinois, United States of America, [1]Department of Biomedical and Biological Sciences, Cornell University, New York, United States of America, [1]Central Department of International Relations and Diplomacy (CDIRD), Kathmandu, Nepal.

Daniel Loebell ORCID iD: 0000-0002-9800-0863
Mingmar Sherpa ORCID iD: 0009-0008-8915-9152
Ram Datta Bhatta ORCID iD: 0000-0002-3680-7282

**ABSTRACT**
The Millennium Challenge Corporation (MCC), initiated in 2004 by the United States (US) Congress, mainly focuses on the initiative for development, which involves good governance, sustainable economic growth, and poverty reduction. Since its establishment, it has invested in 30 countries, accounting for more than $13 billion. Nepal has been one of the recent beneficiaries of the MCC, signing the compact valued at $500 million with the US in 2017, which was later ratified in 2022. The compact has invested mainly in two sectors of development, road infrastructure and electricity transmission. The compact project involves the construction of 315 km of high voltage transmission lines, three substations, and the upgrade of 100 km of the East-West highway. Supporting the commercialization of Nepal's 40 GW hydropower potential, the MCC compact helps in addressing the structural economic prosperity of Nepal. In addition to the economic contribution, the compact also has an impact on the broader implications in Nepal's foreign policy. The MCC compact exemplifies the strategic shift of diversifying the partnership in development, reducing the overdependence on its neighbors, China and India. Similarly, the establishment of the Millennium Challenge Account (MCA)-Nepal gives the sense of host-country ownership to Nepal, reducing the concern over geopolitical influences. Our paper builds on International Relations scholar Shiping Tang's Institutional Foundations for Economic Development (IFED) framework by demonstrating that the MCC fills a crucial area for Nepal's capacity for infrastructure development. Success stories, as demonstrated in Ghana and other countries under the MCC compact, where the compact put forward a transformative step in the pursuit of prosperity, Nepal has leverage in both development and diplomatic paths. Drawing on various articles, this paper argues that the MCC compact has provided Nepal with the capacity for possibilities that will help the nation overcome infrastructural barriers to growth.

**Keywords:** Millennium Challenge Corporation, sustainable development, economic development, transportation, infrastructure.

**INTRODUCTION**
The Millennium Challenge Corporation (MCC) was established in January 2004 by the Congress of the United States (US), which is termed as an independent and innovative foreign assistance agency of the US, aiming at the reduction of global poverty through economic growth (Millennium

Challenge Corporation, 2021). The MCC is perceived as the most substantial change in the foreign aid policy of the US (Phuyel, 2025). In contrast to traditional aid organizations, the MCC is targeted specifically to the governments of those countries that are committed to upholding effective governance by means of the rule of law, fostering economic independence, and respecting democratic rights of their citizens as determined by a set of transparent and objective governance indicators (Phuyel, 2025). This discussion forms the basis of Nepal's effort to assert its independence and maintain mutual respect as a sovereign state while strategically balancing its relationship with India, China, and the US in the pursuit of economic development.

In September 2017, Nepal officially signed the $500 million compact with the US government, with an independent contribution of additional $130 million (Adhikari, 2022). The MCC supports the construction of 315 km of 400 kV transmission line from Lapsiphedi to New Butwal with three associated substations, and the 77 km road section of the East-West Highway in Dang District will be upgraded with new technology (Shrestha, 2023). Nepal signed the MCC implementation agreement in 2019. Therefore, the MCC has two important agreements: first, the MCC agreement, also called the compact, and second, the program implementation agreement, which is termed as the implementation agreement. The ultimate goal is to strengthen Nepal's energy sector by building a robust transmission line, increasing electricity availability, and facilitating cross-border power trade with India (Nepal, 2025).

According to the United Nations Economic and Social Commission for Asia and the Pacific (UNESCAP), funds provide by government budgets in the infrastructure investments in countries with Special Needs (CSN) is 65%, funds by the private sector is 15%, loans and credits from Multilateral Development Banks (MDBs) is 10%, and by Official Development Assistance (ODA) is 10%. Infrastructure investments in developing countries are publicly funded by the budget by just 30% (Mahat, 2021). Given the substantial funding gap in infrastructure, all financing sources, including state, private, domestic, and international, would be needed to meet their infrastructure requirements. These measures show that the US government's MCC compact can help Nepal in both theoretical and practical contexts in addressing its enduring economic challenges, which is crucial for the nation's prosperity.

**METHODS**
This article is written based on the study conducted on available literature, news reports, publicly available government databases, and official MCC and MCA websites. All literature and news reports were selected based on relevance to our topic of interest, credibility, and contribution to the subject. Primary and official databases were given priority, where possible, for the enhancement of reliability.

**NEPAL MCC INITIATION**
The concise agreement of the MCC was concluded in 2017 following extensive discussions with the governmental bodies, enterprises, and civil society organizations. The emphasis was on constructing fundamental infrastructure, particularly for road maintenance and electrical transmission, to circumvent legally binding constraints on economic growth (Millennium Challenge Corporation, 2017b). Nepal's Federal Parliament ratified the MCC in 2022, with making intensive political discourse and public protests. Legislators ultimately ratified the accord, despite forceful protest from leftist parties and civil society organizations, which believes it as a part of the US geopolitical strategy in South Asia. Two leftist parties, CPN-MC and CPN-US, pressured their parliamentarians to vote in support of the MCC compact in an unexpected twist, after an agreement to approve the Twelve Point Interpretative Declaration (Maskey, 2024).

The ratification of the MCC compact has precipitated a strategic transformation in Nepal's foreign aid diplomacy. It illustrates how pragmatic international involvement is supplementing conventional nonalignment, as external development partnerships are utilized to obtain assistance and strategic autonomy.

The aid coming from the MCC program has not been discussed from an economic development perspective amid weak economic indicators in the country. The possibilities of employment, entrepreneurial activities, state of transportation, and shifts in electricity pricing and transport facilities will see imminent positive results after the implementation of the MCC projects (Adhikari, 2022). The shift in the economic development perspective has shadowed opportunities related to employment and poverty alleviation. It signifies a minor yet major change in our perspective, transitioning from dependence on minimal assistance to collaborating with individuals worldwide for self-sufficiency.

## MCC REGULATION: MCA-NEPAL

The MCA-Nepal, owned by the government of Nepal, was created after 2017 to ensure the host country's ownership. MCA-Nepal has the obligation to ensure the project management, procurement, and compliance with the legislation of Nepal. This regulation allays the concerns over foreign domination by providing a practical application of Kantian ethics, where autonomy and mutual benefit drive international collaboration. Recent sources claim that when the US funding freeze was lifted in July 2025, MCA-Nepal resumed full-fledged operations, speeding up stalled procurement processes.

## THE IMPACT OF THE MCC COMPACT ON NEPAL'S ENERGY AND TRANSPORTATION INFRASTRUCTURE DEVELOPMENT

In Nepal, the MCC compact has an investment totaling $697 million ($500 million from the US government + $197 million from the government of Nepal) (Millennium Challenge Corporation, 2017b). The two main areas of investment will be on developing an electricity transmission project, where the project aims to develop 315Km of 400KV transmission lines and commercializing the 40 GW hydropower potential in the energy industry. Second is the Road Maintenance Project (RMP) that will deliver cleaner and sustainable road infrastructure that involves International Road Assessment Programme safety standards (Millennium Challenge Corporation, 2017b). Figure 1 shows the breakdown of budgeting for the MCC compact (Millennium Challenge Corporation, 2017b). With MCC's expansion, many underserved regions of Nepal will be directly benefiting with transformative step towards sustainable development of the country.

It is estimated that Nepal has a hydropower potential of 83,000 MW, with around 43,000 MW being practical and financially affordable (Bhatt & Joshi, 2024). But so far Nepal has only out produced around 5% of what it can produce practically (Bhatt & Joshi, 2024). Until March 2022, Nepal had a total power capacity of 2,205 MW, of which 2,022 MW came from hydropower, 49.76 MW from solar plants, 52.4 MW from thermal plants, 6 MW from sugar mills through cogeneration technology, and 74 MW from alternative energy promotion centers (Ministry of Finance, 2022). Around 90% of energy production in Nepal depends on hydropower alone, which shows its importance. In mid-March 2022, 94% of the country's population had access to electricity, whereas still 6% of the population is out of reach from electricity (Ministry of Finance, 2022).

This MCC project is projecting the development of high high-capacity 400KV transmission line and three new substations that directly provide benefits to 23 million Nepalese people (Millennium Challenge Corporation, 2017b). Similarly, the project will develop 315 Km of new transmission lines, which will serve as the backbone of the country's infrastructure and energy future (Millennium Challenge Corporation, 2017b). Such capacity building not only builds infrastructure but also ensures the accessibility, affordability, and reliability of electricity for Nepal. For Nepalese people, such infrastructure building means reliable energy access, which means quality of life and support to build the potential of entrepreneurship. Figure 2 shows how an increase in electricity per capita increases the human development index (Attigah & Mayer-Tasch, 2013). Businesses can thrive more with the consistent supply of power, which can contribute to economic growth and job creation in the country. Mitigating the scarcity of energy drives economic growth (Stern et al., 2019). Currently, Nepal imports 654 MW of electricity from India, including 54 MW from the Tanakpur-Mahendranagar transmission line and 600 MW from the 400 KV Dhalkebar-Muzaffarpur transmission line (Forum, 2025). At present, there are 228 hydropower projects under construction in Nepal as of May 2023, which will generate 8,434 MW of electricity (Investopaper, 2025). and with the completion of this MCC compact project, Nepal will benefit from export opportunities to neighboring countries, which will support the economic growth of the country. Moreover, it will reduce the burden of being dependent on fossil fuel for daily uses such as cooking, heating, etc., which also has an environmental benefit for the country where climate change has been the long-standing national challenge.

Transportation infrastructure is taken as the way to promote the county's economic prosperity and growth (Banerjee et al., 2020). An accessible, reliable, and affordable infrastructure network is important for the development of the nation (Luo & Xu, 2018). Nepal has seen a significant growth in its road infrastructure in the past decades, connecting many parts of the country. In 2021, Nepal's road network reached a stunning 80,078 Km, with blacktopped roads being 15,935 Km, 20,082 Km graveled, and 44,061 Km unpaved (Department of Roads). The country's geography, limited budget, and natural disasters are some of the challenges that Nepal faces when it comes to building road infrastructure. Similarly, road safety is yet another pressing issue in Nepal, mainly due to poor road conditions and ineffective traffic management (Bhattarai & Shafir, 2025).

The RMP under MCC includes periodic maintenance of 90 Km of Nepal's strategic road infrastructure, and 40 Km of Full-Depth Reclamation (FDR) by recycling the existing pavement material to create new roads base reducing the cost of maintenance (Millennium Challenge Corporation, 2017b). The project also provides technical assistance programs to different institutions and agencies that are focused on road maintenance (Millennium Challenge Corporation, 2017b). Such investment will facilitate the efficient movement of goods and people, which reduces the cost of transportation and operating costs. Similarly, in turn, it will encourage investment and support rural development with the connectivity of good roads. The RMP will work in collaboration with the Department of Roads, the Road Board Nepal, and the Institute of

Engineering at Tribhuvan University on training, provision of equipment, and knowledge transfer that can help Nepalese engineers to carry out the RMP's investments in the future (Millennium Challenge Corporation, 2017b). Moreover, the collaboration will support to develop skilled manpower for the country for future infrastructure initiatives.

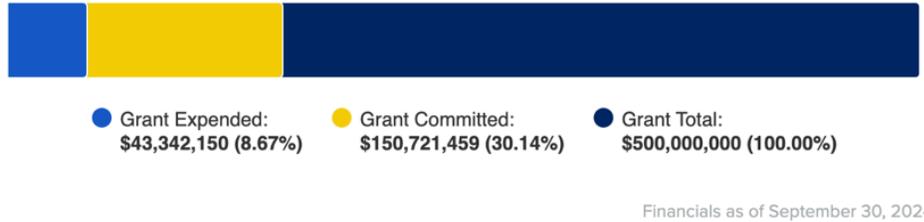

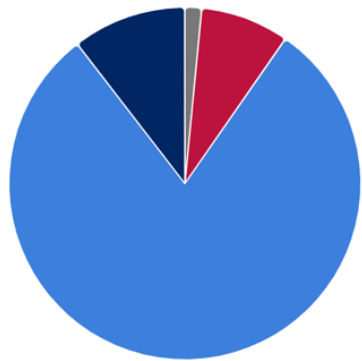

**Figure 1:** MCC-Nepal Compact (Millennium Challenge Corporation, 2017b).

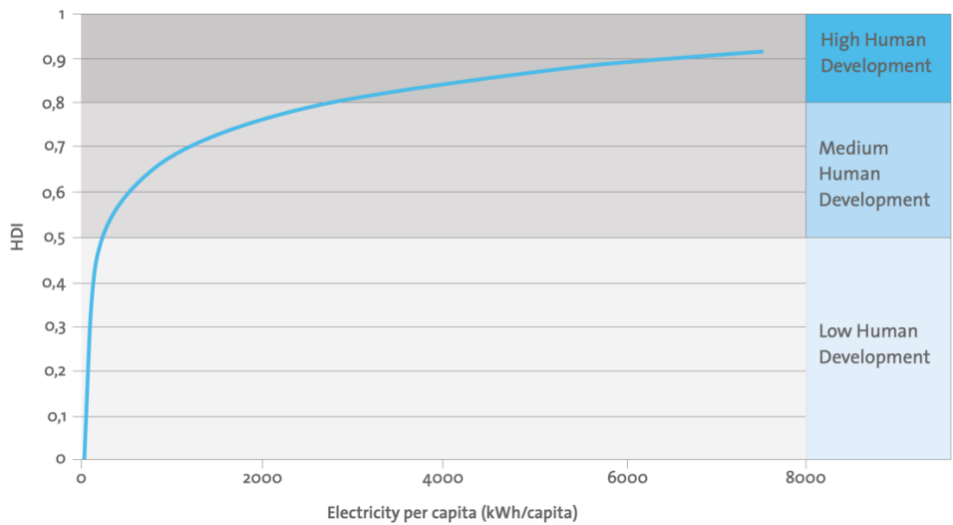

**Figure 2:** Correlation between electricity and Human Development Index (Attigah & Mayer-Tasch, 2013).

**TRANSFORMATIVE IMPACT OF THE MCC IN THE WORLD**
The MCC projects have significantly contributed to economic growth, poverty reduction, and good governance. One of many examples is the success in Ghana, where two MCC compacts have significantly transformed the agriculture, road, and energy sectors of the country. The first MCC compact for Ghana (2007-2012), worth $547 million, funded the project such as: $228 million in transportation projects, which helped in upgrading 14 Km of the N1 highway in Accra and 75 Km of the Agogo-Dome Trunk Road, and constructed two ferries in the Afram Basin Zone (Millennium Challenge Corporation, 2017a). The second MCC compact (2016-2021) invested $498 million, which aimed at reforming the energy sector of the country by expanding access to electricity in underserved areas, reducing power losses, and supporting utility improvements (Millennium Challenge Corporation, 2020). Ghana, as an emerging economy and a trade center for most of West Africa, expansion of transportation project supported export directed horticulture base by minimizing the cost of transportation for agricultural products. The Ghana compact was successful in creating 1,200 farmer-based organization, 66,930 farmers were trained in commercial agriculture, enhanced rural credit and banking services, and constructed new roads and post-harvest facilities (Millennium Challenge Corporation, 2017a). Similarly, the MCC-Ghana power compact has increased the country's transmission network capacity by 1,015 MVA, which represents 10% of Ghana's transmission capacity (Millennium Challenge Corporation, 2017a). Moreover, from the Power Compact, more than 300 female students in the field of Science, Technology, Engineering and Math (STEM) found internship positions as some of the renowned energy institutions, and more than 600 female STEM students also participated in mentoring and training for professional growth and development (Millennium Challenge Corporation, 2017a).

Similarly, the MCC has a larger impact in other regions, including Mongolia, where it funded infrastructure projects such as building roads and wastewater treatment plants (Millennium Challenge Corporation, 2013). In El Salvador, the project enhanced rural electrification and skills training (Millennium Challenge Corporation, 2012). Finally, the model of result-based financing makes the MCC projects significantly impactful, which provides long-term economic growth for countries. In the world, the MCC has been a trusted development partners, which empowers people, demonstrating its success in fighting against poverty.

**REFLECTING ON THE LITERATURE OF INSTITUTIONAL FOUNDATIONS OF ECONOMIC DEVELOPMENT (IFED)**
Much of the literature that we draw from is described in the first chapter of Shiping Tang's book on the institutional foundations of economic development (IFED) (Tang, 2022). So, as not to tread over much of the same ground, this section will summarize the main works that Tang focuses on. He states that scholars representing two different bodies of economics literature have both failed to sufficiently identify institutional elements for development, because they have focused solely on inductive analyses and have not utilized deduction to create a balanced, systematic statement on IFED.

The first body is the old institutional economics literature, which is best identified by Adam Smith. In a lost 1755 paper that was later quoted by his student, Dugald Stewart, Smith explained that "little else is requisite to carry a state to the highest degree of opulence from the lowest barbarism, but peace, easy taxes, and a tolerable administration of justice; all the rest being brought about by the natural course of things" (Stewart & Edinburgh, 1794, p. 72). In this passage, Smith indicated fair institutions for taxation and adjudication. Two centuries later, the black economist William Arthur Lewis elaborated on the purpose of institutions by positing their three roles: protecting the efforts of economic actors, providing opportunities for specialization, and permitting freedom of maneuver (Lewis, 1955, p. 57).

The second body is the new institutional economics literature, best known for economists who focus on property rights and transaction costs like Douglas North, Elinor Ostrom, and Amartya Sen. North and his co-authors explained that two important institutions that impact development are property rights and incentive structures that are underpinned by exemption from expropriation by the state and individuals (North & Thomas, 1973), (North & Weingast, 1989). The issue with these inductive analyses were that, as Tang explains, they were singularly inspired by the British experience in that they claim property rights are the key because Britain became the first industrialized society through institutionalizing property rights, constraints on executives in government, and credible commitments to protecting rights (Andrews, 2023, p. 27). Elinor Ostrom identified seven types of rules for public choice that stimulate development: governing position, boundary, scope, authority, aggregation (in decision making), information, and payoff (Ostrom, 1986). However, she did not provide any explanation for why these types were so crucial for stimulating growth. Amartya Sen differentiates two types of freedoms that institutions can provide: freedom of action and decision as well as actual opportunity. However, he then stretches the concept of freedom by providing five types – political, economic facilitative, social opportunities, transparency guarantee, protection security – without any strong rationale for doing this (Sen, 2000, pp. 17-20).

Tang addresses these deficiencies by contending that institutions (and their policies) shape development through six dimensions that govern four immediate social drivers of individuals actions:

| Drivers | Institutional dimensions |
|---|---|
| Possibility | 1. Hierarchy for order and stability<br>2. Liberty for protecting innovation (as possibilities) |
| Incentive | 1. Property rights and beyond: for incentives in the material market<br>2. The channels of social mobility: for incentives in the positional market |
| Capability | Redistribution that impacts capabilities, directly and indirectly |
| Opportunity | Equality of opportunity, or affirmative action for employment and bidding |

**Table 1:** The Four Immediate Drivers and the Six Dimensions (Tang, 2022, p. 40).

Hence, institutions underpin four immediate drivers of productive activities when it comes to development, and incentives is only one of them. The other three roles are to govern individuals' and organizations' possibility, capability, and opportunity for taking productive initiatives. Without incentives, individuals will not make an effort even if they have possibility and capability. Without possibility and capability, however individuals cannot make an effort even if there are incentives offered. Without equality of opportunity, some individuals will be excluded from taking initiatives. By single-mindedly focusing on institutions' role of governing (material) incentives, NIE [new institutional economics] has been neglecting the other three functions or roles of institutions almost completely.

From the four immediate drivers of individual actions or initiatives, we can then deduce that six dimensions of institutions should underpin the four drivers, and these six dimensions form IFED. In other words, according to our framework, six dimensions of institutions shape economic development by impacting four drivers of individuals' actions and initiatives, which are the ultimate drivers of economic development (Tang, 2022, p. 40).

Here, Tang offers crucial insight about the IFED. However, he ignores a crucial dimension that, even when developing countries are on their way to applying all of the other six dimensions that he describes, they would be severely restricted in their prospects for economic development. Nepal, as a state that has seen almost 20 years of peace and democratic transition, offers important insight into how a state can offer possibility, incentives, and other aspects but still come up short when it faces natural barriers to infrastructural development and, afterwards, encouraging the kind of economic incentives he describes.

We have shown how another element of possibility needs to be added and prioritized above all other dimensions to account for development. Most notably, under Tang's element of "possibility" there needs to be "capacity for possibility." Additionally, the ways in which countries can achieve such capacity, if they do not have the financial resources to do so themselves, need to be through due diligence in selecting foreign investment. For this reason, our research suggests that the US government's MCC provides the means for Nepal to achieve sufficient capacity to catalyze Tang's six dimensions. We make this recommendation over other sources of investment (namely those originating in India and China) because, unlike the latter two, the MCC has a proven track record of helping other countries achieve this kind of capacity.

**CONCLUSION**
Nepal's involvement in the MCC involves the pragmatic philosophy of Nepal's foreign policy of sovereignty and global interdependence nurture on its economic necessity. The infrastructure component of the agreement promises to spur prosperity in line with Nepal's foreign policy goals while negotiating geopolitical sensitivities with China, India, and the US. Success of the MCC compact in Nepal depends on the transparency of the MCA-Nepal governance, timely implementation, which also present the clear balance between aid and autonomy. This partnership demonstrates Nepal's ability to improve its economic circumstances by overcoming both external and internal factors.